\documentclass[12pt,preprint]{aastex} 

\usepackage{natbib}

\begin{document}

\title{Magnetic Modulation of Stellar Angular Momentum Loss}

\author{ Cecilia Garraffo\altaffilmark{1}, Jeremy
  J. Drake\altaffilmark{1}, and Ofer Cohen\altaffilmark{1}}

\altaffiltext{1}{Harvard-Smithsonian Center for Astrophysics, 60 Garden St. Cambridge, MA 02138}

\begin{abstract}

Angular Momentum Loss is important for understanding
astrophysical phenomena such as stellar rotation, magnetic
activity, close binaries, and cataclysmic variables. Magnetic breaking
is the dominant mechanism in the spin down of young late-type
stars. We have studied angular momentum loss as a function of stellar magnetic
activity. We argue that the complexity of the field and its
latitudinal distribution are crucial for angular momentum loss
rates. In this work we discuss how angular momentum is modulated by
magnetic cycles, and how stellar spin
down is not just a simple function of large scale magnetic field
strength.

\end{abstract}

\section{INTRODUCTION}

Stars spin-down through their magnetized winds that carry away mass and
angular momentum (magnetic breaking). These winds
corotate with them up to a certain distance, called the Alfv\'en radius ($R_A$), where the speed of the wind
reaches the Alfv\'enic speed ($ u_A=B/\sqrt{4\pi \rho}$, being $B$
the magnetic field strength and $\rho$ the density of the plasma).  The
strength of the wind is expected to scale up with field strength and, therefore, stellar spin-down is a function
of magnetic activity.  \citet{1967ApJ...148..217W} provided the first analytical expression for
magnetic breaking, $\dot{J}=\frac{2}{3} \Omega \dot{M}R_A^2 $, where
$\dot{J}$ is the angular momentum loss rate, $\Omega$ the
angular velocity, $\dot{M}$ the mass loss rate, and where spherical symmetry has been
assumed (a split magnetic monopole was used and $R_A$ is
constant).  

Recently, improved observational techniques like Zeeman
Doppler Imaging \citep{1997A&A...326.1135D} have provided us with new information about the magnetic
topology of almost a hundred stars \citep[see][and references
therein]{2009ARA&A..47..333D, 2010MNRAS.407.2269M}.  The fact that most of them show much more
complex structure than that of a dipole
has drawn a great amount of attention
towards the role of magnetic topology in angular momentum loss.  

Since then, some authors have discussed the importance of topology for
the winds structure and efficiency of magnetic braking
 \citep{2009ApJ...699.1501C, 2013ApJ...764...32G, 2014MNRAS.439.2122L, 2011ApJ...741...54C}.  
Both analytical and numerical models have been built aiming, with increasing
success, to realistically study the interplay between spin-down
rates and magnetic structure at the surface of stars
\citep{1987MNRAS.226...57M, 1988ApJ...333..236K, 1995ApJ...446..435C, 1958ApJ...128..677P,
  2012ApJ...754L..26M, 2011ApJ...741...54C, 2009ApJ...699.1501C, 2014ApJ...783...55C}. 

In this work we study the relevance of magnetic active regions and their
latitudinal locations on stellar spin-down rates. 
Using a three dimensional magnetohydrodynamic (MHD) code we perform simulations to
explore the effect of stellar spots on mass and angular momentum loss rates.

\section{NUMERICAL SIMULATION}

We use the generic {\it BATS-R-US} code \citep{1999JCoPh.154..284P,2012JCoPh.231..870T} that solves the set of MHD equations for the conservation of mass, momentum, magnetic induction, and energy on a spherical, logarithmic in the $\hat{r}$ coordinate grid. We take advantage of the code's Adaptive Mesh Refinement capabilities to refine the grid around regions where the magnetic field changes its sign. This way we better resolve current sheets that form during the simulation.

We carry out simulations for a set of magnetic maps, starting with two pure
dipoles (with field strengths of 10 and 20 Gauss at the poles) and
superimposing the magnetic active regions observed for the sun near solar maximum
(Carrington Rotation 1958). We place these two rings of spots at three different latitudes, using the
same shift in latitude for both hemispheres, and with three different
scalings for the spots strength. 

From the resulting solutions we compute the mass and angular momentum
loss rates and study the spatial distribution of the
contributing quantities (density and wind speed) at the Alfv\'en and stellar
surfaces, as well as the shape and size of the Alfv\'en Surface
itself.

\section{RESULTS AND DISCUSSION}

\begin{figure}[ht!]
\centering
\includegraphics[width=\textwidth]{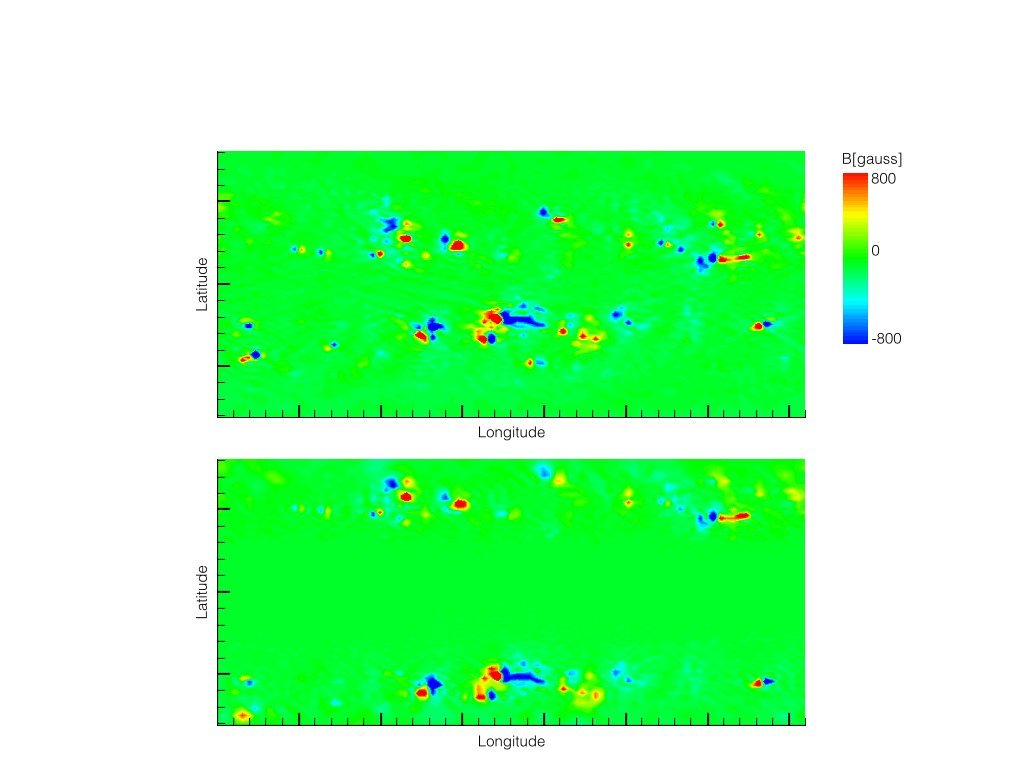}
\caption{Typical stellar magnetograms used in our simulations for magnetic active regions at low (top) and high (bottom) latitudes}
\label{fig:1}
\end{figure}

The general behaviour is ruled by the latitude of the magnetic active
regions and is well described by two limiting cases: magnetograms with
low-latitude and high-latitude spots (see figure \ref{fig:1}).  We group our results in these two categories that represent the general trend of
the whole sample.

\begin{figure}[ht!]
\centering
\includegraphics[width=\textwidth]{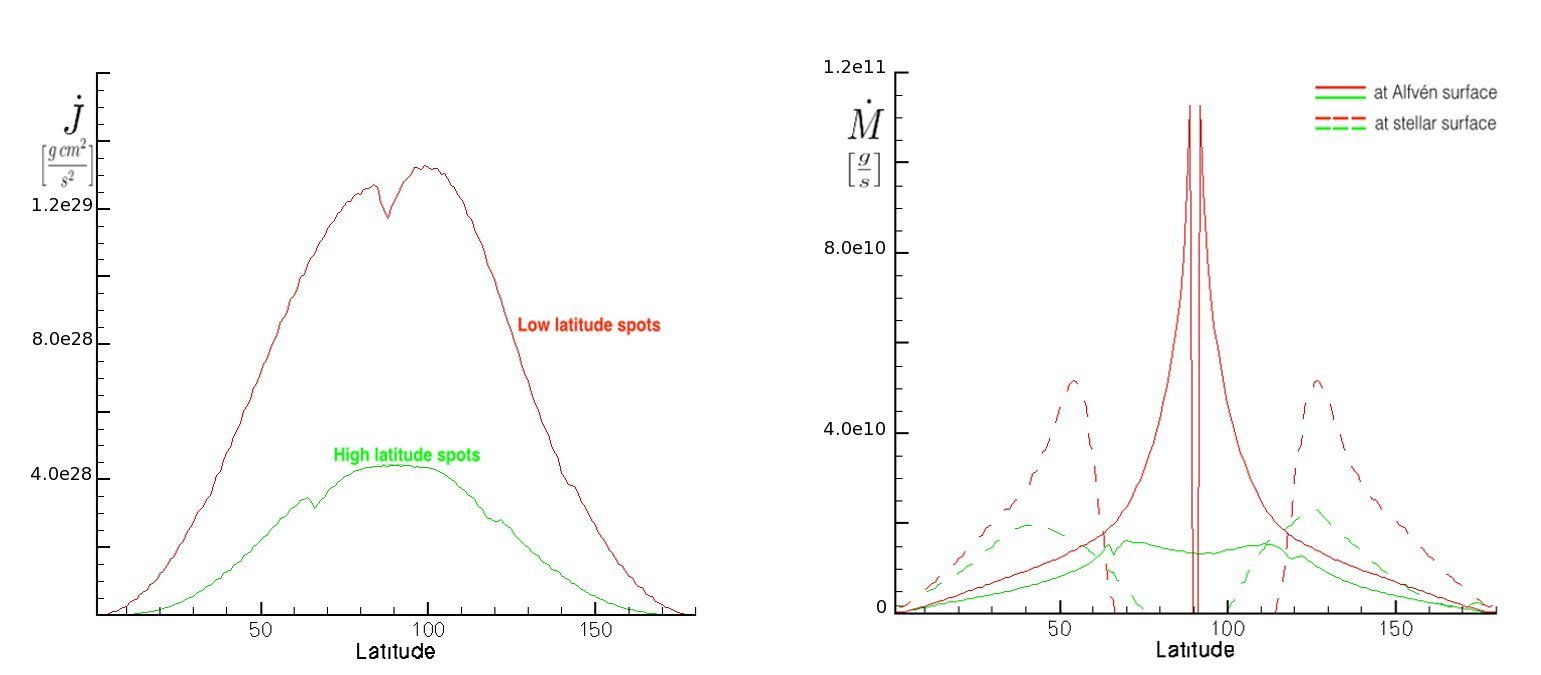}
\caption{Mass loss and angular momentum loss as a function of latitude for low-latitude (red) and high-latitude (green) magnetic active regions at the Alfv\'en Surface (solid line) and at the stellar surface (dashed line).}
\label{fig:2}
\end{figure}

In order to illustrate the qualitative effect of the magnetic active
regions in each of these two regimes, in Fig~\ref{fig:2} we
compare the mass (both at the Alfv\'en and at the stellar surfaces) and angular momentum losses as a
function of latitude for a magnetogram with low-latitude spots and the same with
the spots shifted towards high-latitudes.  The transition latitude from
one regime to the other is determined by the limiting latitude between
open and closed field lines regions on the stellar surface of the pure
dipolar solution.

We find that, for all cases, there is a significant reduction of mass
and angular momentum loss when magnetic active regions are located at
high-latitudes compared to both the pure dipolar solution and the one
with low-latitude
spots.

\begin{figure}[ht!]
\centering
\includegraphics[width=\textwidth]{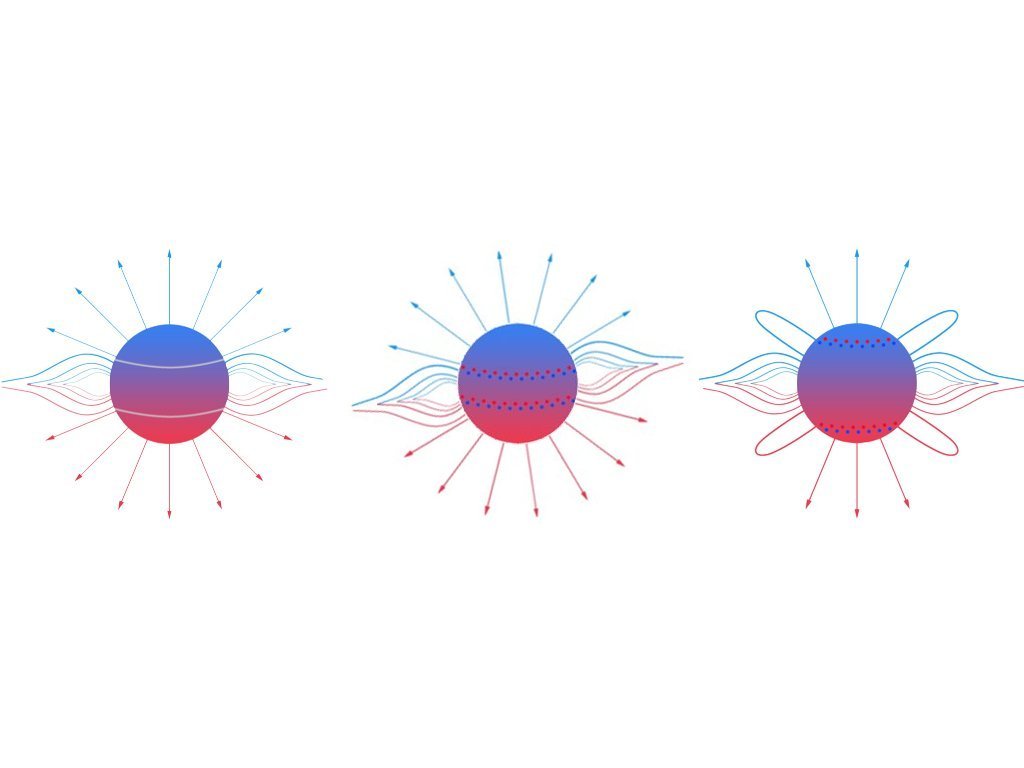}
\caption{Qualitative plot of the wind structure for a dipole (left),
  with the limiting latitude between open and closed field line
  regions shown in white.  The wind structure for the same dipole with
  low-latitude (center) and high-latitude (right) magnetic active
  regions.  The addition of spots at fairly low latitudes---within the
  "dead zone" where field is already closed---makes little difference
  to the wind morphology or mass and angular momentum loss rates, and
  only results in a tilt of the magnetic axis.  Adding spots at higher latitudes closes field lines and quenches mass and angular momentum loss.}
\label{fig:3}
\end{figure}

In general, the introduction of magnetic active regions within the
``dead-zone'' from which wind does not escape for the dipolar case
(low-latitudes) does not change systematically the
mass or angular momentum loss rates.  However, any reallistic
distribution of active regions will produce a small residual dipolar
component as a result of the spatial separation of the spots. For
rings of spots at low-latitudes, this competing dipole will be almost
perpendicular to the rotation axis and original dipole, and will result
in a tilt of the Alfv\'en surface.  Consequently, the low-latitude
active regions lead to a change in orientation of the Alfv\'en
surface.  
In contrast, the introduction of active regions at high-latitudes leads to a
significant reduction in both mass and angular momentum loss rates.
The reason for this is that spots located in the open field lines
regions of a star can efficiently couple to them and close them, therefore reducing the amount
of wind carrying plasma away (see Fig~\ref{fig:3} for a qualitative plot). 

It is also worth noticing from Fig~\ref{fig:2} that most of the mass being lost is
coming from mid-latitudes at the stellar surface and it funnels
towards the equator when approaching the Alfv\'en surface.  This makes
sense because closed field lines cannot contribute to the mass loss
mechanism.  As a result, most of the torque experienced by the star
will be located in a narrow band at mid latitudes, fundamentally at the transition
latitude where the open field lines begin to appear and that that divides the low and high-latitude regimes.   

In summary, we find a bi-modal regime regulated by the latitude of spots by which magnetic
active regions efficiently reduce mass and angular momentum
loss rates {\it only} when located outside of the dead zone (the closed field line
region on the stellar
surface).  The mechanism behind it is the closing of otherwise open
field that leads to a reduction in the plasma being carried
away (see Fig~\ref{fig:3}).  As a result, magnetic cycles of stars whose spots cross
the limiting latitude of the dead-zone, and therefore turn on and off this
mechanism,  will experience a large modulation of mass
and angular momentum loss, by a typical factor of the order of a few (see
Fig~\ref{fig:2}).  

\newpage

\acknowledgments

We thank Jean-Ren\'e Galarneau for providing Figure~\ref{fig:3}.  CG
was supported by {\it Smithsonian Institution Consortium for Unlocking
  the Mysteries of the Universe} grant'' Lessons from Mars: Are Habitable Atmospheres on Planets around M Dwarfs Viable?'' during the course of this research. 
JJD was supported by NASA contract NAS8-03060 to the {\it Chandra X-ray
Center} (CXC) and thanks the CXC director, Belinda Wilkes, and the CXC
science team for advice and support. 
OC was supported by the {\it Smithsonian Institution Consortium for
  Unlocking the Mysteries of the Universe} grant ''Lessons from Mars: Are
Habitable Atmospheres on Planets around M Dwarfs Viable?'' and by the
{\it Smithsonian Institute Competitive Grants Program for Science} (CGPS)
grant ''Can Exoplanets Around Red Dwarfs Maintain Habitable
Atmospheres?.''
The simulations were performed on the NASA HEC Pleiades system under award SMD-13-4526.

\end{document}